\documentclass{JHEP3}
\usepackage{amsmath,amssymb}

\newcommand{\com}[2]{[#1,#2]}
\newcommand{\starcom}[2]{[#1\stackrel{\star}{,}#2]}
\newcommand{\pair}[2]{\langle #1,#2\rangle}
\newcommand{\Nabla}{\triangledown}

\title{%
  Cosmological and Black~Hole Spacetimes in Twisted~Noncommutative~Gravity}

\author{%
  Thorsten~Ohl and Alexander~Schenkel\\
  Institut f\"ur Theoretische Physik und Astrophysik, Universit\"at W\"urzburg, \\
  Am Hubland, 97074 W\"urzburg, Germany\\
  E-mail: \email{ohl@physik.uni-wuerzburg.de}, \email{aschenkel@physik.uni-wuerzburg.de}}

\abstract{%
  We derive noncommutative Einstein equations for abelian twists and
  their solutions in consistently symmetry reduced sectors,
  corresponding to twisted FRW cosmology and Schwarzschild black
  holes.  While some of these solutions must be rejected as models for
  physical spacetimes because they contradict observations, we find
  also solutions that can be made compatible with low energy phenomenology,
  while exhibiting strong noncommutativity at very short distances and
  early times.}


\preprint{\today}

\begin{document}

\section{Introduction}
Despite the great success of Einstein's general theory of relativity,
it is generally believed that it has to be modified at small
distances, incorporating quantum effects of spacetime. To achieve this
goal and arrive at a consistent theory of quantum gravity, a number of
different approaches have been proposed, including string theory and
loop quantum gravity as prominent examples.  The aim of these
models is to provide a microscopic description of quantum spacetime
subsequently to make contact to more macroscopic phenomena, like
e.\,g.~our universe. In doing so it turns out that it is quite hard to
connect the very small length scales on which these models are defined
with the large scales on which observable physics takes
place, e.\,g.~cosmic inflation or particle physics.

A complementary approach towards quantum gravity is to construct
effective theories as an intermediate step between general relativity
and a full theory of quantum gravity and study physical applications
within it. These results can then possibly be used to connect full
quantum gravities to physical phenomena. There have been many
approaches in this direction, from which we choose the approach of
noncommutative (NC) gravity based on a modification of
symmetries~\cite{Aschieri:2005yw,Aschieri:2005zs} (see also the
review~\cite{Aschieri:2006kc}). The main idea behind this formalism is
to replace the classical symmetries of general relativity (i.\,e.~the
diffeomorphisms) by a twist deformed Hopf algebra of diffeomorphisms,
which can be interpreted as quantum symmetries. As a result of this
postulate one obtains that these theories naturally live on
noncommutative spacetimes.

For effective theories, it is crucial to find solutions and study the
physics they describe.  For the case of the gravity theory proposed
in~\cite{Aschieri:2005yw,Aschieri:2005zs} the only solutions known are
the NC black hole models discussed in~\cite{schuppbh}. This lack of
models provided the motivation for us to systematically discuss
symmetry reduction in Hopf algebra based gravity theories in our
previous paper~\cite{Ohl:2008tw}. This can be seen as a first step
towards the construction of solutions.  We found that there are
consistency conditions restricting compatible Drinfel'd twists for a
given symmetry. This has lead to a classification of admissible
deformations of Friedmann-Robertson-Walker universes and Schwarzschild
black holes in the presence of a certain type of twist deformations,
the so-called abelian or Reshetikhin-Jambor-Sykora~(RJS)
twists~\cite{Reshetikhin:1990ep,Jambor:2004kc}.

The main goal of this brief paper is to show that most of our models
solve the noncommutative Einstein equations proposed
in~\cite{Aschieri:2005zs}. Furthermore, some physical implications of
our models are discussed.  Therefore, we briefly review the models
proposed in~\cite{Ohl:2008tw} in section~\ref{sec:models} and discuss
the phenomenologically interesting models. In section~\ref{sec:basics}
we briefly review the noncommutative Riemannian geometry and Einstein
equations of~\cite{Aschieri:2005zs}.  We will give a simplified
formalism for the case of RJS twists by using a special basis of
vector fields and one-forms on the manifold. In
section~\ref{sec:cosmo} and~\ref{sec:blackhole} we show that most of
our models solve the noncommutative Einstein equations and discuss
some phenomenology. In section~\ref{sec:conc} we conclude and give an
outlook to possible future investigations in this field.

\section{Review of Our Cosmological and Black Hole Models}
\label{sec:models}
In our previous paper~\cite{Ohl:2008tw} we have classified possible noncommutative 
Friedmann-Robertson-Walker (FRW) cosmologies and Schwarzschild black holes in the presence of a
 Reshetikhin-Jambor-Sykora (RJS) twist~\cite{Reshetikhin:1990ep,Jambor:2004kc}.
 These twists are given by
\begin{flalign}
 \label{eqn:RJStwist}
\mathcal{F}_V := \exp \Bigl(-\frac{i\lambda}{2} \Theta^{\alpha\beta}V_\alpha\otimes V_\beta\Bigr)~,
\end{flalign}
where $\lbrace V_\alpha\in \Xi\rbrace$ is a set of mutually commuting vector fields and $\Theta^{\alpha\beta}$ can
 be taken in the canonical (i.\,e.~Darboux) form.

Using the $\star$-commutators $\starcom{x^\mu}{x^\nu}$ among the
linear coordinate functions from the appendix of~\cite{Ohl:2008tw}, we
can restrict our models to physically sensible cases.  As a criterion
for the cosmological models $\mathfrak{C}_{AB}$ we demand that the
scale of noncommutativity does not grow in physical length
scales. This excludes in particular the Moyal-Weyl type model
$\mathfrak{C}_{11}$ with $\starcom{x^i}{x^j}=i \theta^{ij}$, where
$\theta^{ij}=\mathrm{const.}$, since $x^i$ are comoving spatial
coordinates and have to be multiplied by the scale factor of the
universe $A(t)$ in order to give physical length scales. Since the
scale factor grows rapidly during inflation, the physical scale of
noncommutativity $A(t)^2 \theta^{ij}$ grows too, leading to a very
noncommutative late universe, which contradicts
observations. Including analogous arguments for the time-space
$\star$-commutators we obtain the physically valid models
$\mathfrak{C}_{22}$ with $\mathbf{c}_2=0$ or $\vert V_1^0(t)
A(t)\vert$ nongrowing in $t$ and $\mathfrak{C}_{32}$ with
$\mathbf{c}_{2}=0$. The vector fields $V_\alpha$ generating these
models are given by
\begin{subequations}
\begin{flalign}
 \mathfrak{C}_{22}:&\qquad V_{1}=V_1^0(t)\partial_t ~,\quad V_2=c_2^i\partial_i+d_2^iL_i+f_2 x^i\partial_i~,\\
 \mathfrak{C}_{32}:&\qquad V_{1}=V_1^0(t)\partial_t+d_1^iL_i ~,\quad V_2=V_2^0(t)\partial_t + f_2 x^i\partial_i~.
\end{flalign}
Here $L_i :=\epsilon_{ijk}x^j\partial_k$ are the generators of
rotations and $\com{V_1^0(t)\partial_t}{V_2^0(t)\partial_t}\equiv0$.
Note that we have switched the labels of the vector fields $V_\alpha$
in model $\mathfrak{C}_{22}$ compared to~\cite{Ohl:2008tw} for later
convenience.  Furthermore, we will restrict ourselves to the case
$\mathbf{c}_2=0$ for the model $\mathfrak{C}_{22}$ for the following
reason: the case $\mathbf{c}_2\neq 0$ requires a rapidly decreasing
$V_1^0(t)$ for an inflationary scenario. Thus the model can be well
approximated by the model $\mathfrak{C}_{22}$ with $d^i_2=f_2=0$,
since the additional terms will be suppressed by $V_1^0(t)$ in
physical coordinates. The resulting model is then simply a time-space
Moyal deformation, which has been discussed elsewhere. We will omit an
explicit discussion of this model for brevity and only note that it
can be described by the methods developed below as well.

As a physicality criterion for our Schwarzschild black hole models $\mathfrak{B}_{AB}$~\cite{Ohl:2008tw} we use the requirement 
$N_1=N_2=0$, since otherwise noncommutativity would grow linearly in time. This leads to the physically viable
model $\mathfrak{B}_{12}$ constructed by the vector fields
\begin{flalign}
 \mathfrak{B}_{12}:\qquad V_1 = c_1^0 \partial_t +\kappa_1d^iL_i~,\quad V_2 =c_2^0(r) \partial_t + \kappa_2 d^i L_i+f_2(r) x^i\partial_i~,
\end{flalign}
\end{subequations}
where $c_1^0$ has to be constant and $r:=\Vert\mathbf{x}\Vert$ is the radial coordinate. 
 Note that the other physically viable model $\mathfrak{B}_{11}$ is already included in the class $\mathfrak{B}_{12}$.

We can understand our models better by choosing without loss of generality $\mathbf{d}_1=\mathbf{d}_2=\mathbf{d}=(0,0,d)$ and 
transforming from carthesian coordinates $x^i$ to spherical coordinates $(r,\zeta,\phi)$.
Then the vector fields read
\begin{subequations}
\label{eqn:models}
\begin{flalign}
\label{eqn:c22}
 &\mathfrak{C}_{22}:\qquad V_1=V_1^0(t)\partial_t~,\quad V_2=d\partial_\phi + f_2 r\partial_r~,\\
\label{eqn:c32}
 &\mathfrak{C}_{32}:\qquad V_1=V_1^0(t)\partial_t+d \partial_\phi~,\quad V_2=V_2^0(t)\partial_t + f_2 r\partial_r~,\\
\label{eqn:b12}
&\mathfrak{B}_{12}:\qquad V_1 = c_1^0 \partial_t +\kappa_1\partial_\phi~,\quad V_2 =c_2^0(r) \partial_t + \kappa_2 \partial_\phi +f(r) \partial_r~.
\end{flalign}
\end{subequations}
Note that we have defined $f(r):=f_2(r) r$ and absorbed the parameter
$d$ into $\kappa_\alpha$ in the model $\mathfrak{B}_{12}$ in order to
simplify the expression.

The $\star$-commutation relations among appropriate coordinate
functions in spherical coordinates are
\begin{subequations}
\label{eqn:commutators}
\begin{align}
\label{eqn:a22}
  \mathfrak{C}_{22}:\; &
    \left\{ \begin{aligned}
               \starcom{t}{\exp i\phi}&=-2\exp i\phi~\sinh\Bigl(\frac{\lambda d}{2} V_1^0(t)\partial_t\Bigr) t\\
               \starcom{t}{r}         &=2 i r ~\sin\Bigl(\frac{\lambda f_2}{2} V_1^0(t)\partial_t\Bigr)t
            \end{aligned} \right. \\
\label{eqn:a32}
  \mathfrak{C}_{32}:\; &
    \left\{ \begin{aligned}
               \starcom{t}{\exp i\phi}&= 2 \exp i\phi~\sinh\Bigl(\frac{\lambda d}{2} V_2^0(t)\partial_t\Bigr) t\\
               \starcom{t}{r}         &=2 i r ~\sin\Bigl(\frac{\lambda f_2}{2} V_1^0(t)\partial_t\Bigr)t\\
               \exp i\phi\star r      &= e^{-\lambda d f_2}~r\star \exp i\phi
    \end{aligned} \right. \\
\label{eqn:a12}
  \mathfrak{B}_{12}:\; &
     \left\{ \begin{aligned}
               \starcom{t}{\exp i\phi} &=\exp i\phi~ \Bigl(2 \sinh\Bigl(\frac{\lambda\kappa_1}{2}\bigl(c_2^0(r)\partial_t
                                          + f(r)\partial_r\bigr)\Bigr) t -\lambda \kappa_2 c_1^0\Bigr)\\
               \starcom{t}{r}          &=i\lambda c_1^0 f(r)~,\\
               \starcom{\exp i\phi}{r} &=-2 \exp i\phi ~\sinh\Bigl(\frac{\lambda\kappa_1}{2}f(r)\partial_r\Bigr)r
     \end{aligned} \right..
\end{align}
\end{subequations}
In particular, our models include time-angle, time-radius and angle-radius noncommutativity for both the FRW cosmologies and
the Schwarzschild black holes. Note further that the $\star$-commutators simplify dramatically 
for the choice $V_\alpha^0(t)=\mathrm{const.}$, $f(r)=r$ and $c_2^0(r)=\mathrm{const.}$. This will be further explained below,
 when we discuss specific examples.

\section{Review of Twisted Noncommutative Einstein Equations}
\label{sec:basics}

In this section we will briefly review the noncommutative Riemannian geometry and Einstein equations constructed in~\cite{{Aschieri:2005zs}} (see also~\cite{Aschieri:2005yw} and~\cite{Aschieri:2006kc}). We will restrict the discussion
 to RJS twists (\ref{eqn:RJStwist}).

Since the formulae in~\cite{Aschieri:2005zs} were constructed in a coordinate and basis independent way, we have
 the freedom to choose a suitable basis for the vector fields $\lbrace e_a\in\Xi:a=0,\dots,3\rbrace$ and one-forms
 $\lbrace \theta^a\in\Omega:a=0,\dots,3\rbrace$. It turns out that the expressions for the geometrical quantities and the Einstein 
equations do simplify drastically, if we can find a basis of vector fields $\lbrace e_a\rbrace$ satisfying
\begin{flalign}
\label{eqn:basisconditions}
 \com{e_a}{e_b}=0~,\quad \com{V_\alpha}{e_a}=0~,
\end{flalign}
for all $a,b,\alpha$.
We call the basis (\ref{eqn:basisconditions}) the natural basis of vector fields and construct the 
basis of one-forms $\lbrace \theta^a\rbrace$ by duality 
\begin{flalign}
\delta_a^b=\pair{e_a}{\theta^b}_\star=\pair{\bar f^\alpha(e_a)}{\bar f_\alpha(\theta^b)}=\pair{e_a}{\theta^b} ~,
\end{flalign}
where $\pair{\cdot}{\cdot}$ is the canonical commutative pairing between vector fields and one-forms and $\bar f^\alpha \otimes \bar f_\alpha = \mathcal{F}_V^{-1}$ is the inverse twist.

The existence of a local (densely defined) natural basis (\ref{eqn:basisconditions}) can be shown 
explicitly for the case of RJS twists (\ref{eqn:RJStwist}), 
assuming that the vector fields $V_\alpha$ are analytical almost everywhere.
 In this brief paper we will omit the general proof and only give the natural basis
for our explicit models (\ref{eqn:models}). It turns out that for
the cosmological models $\mathfrak{C}_{22}$ and $\mathfrak{C}_{32}$ we can make the choice
\begin{subequations}
\label{eqn:basiscosmo}
\begin{flalign}
 e_0=v(t)\partial_t~,~~e_1=r\partial_r~,~~e_2=\partial_\zeta~,~~e_3=\partial_\phi~,~~\text{where}\\
v(t)=\begin{cases}
      V_1^0(t) & \text{,~for model~}\mathfrak{C}_{22} \text{~and }\mathfrak{C}_{32}\text{~with~}V_1^0(t)\not\equiv0  \\
      V_2^0(t) & \text{,~for model~}\mathfrak{C}_{32}\text{~with~}V_1^0(t)\equiv0,~V_2^0(t)\not\equiv0\\
      1        & \text{,~for model~}\mathfrak{C}_{32}\text{~with~}V_1^0(t)\equiv V_2^0(t)\equiv0~.
     \end{cases}
\end{flalign}
\end{subequations}
For the black hole model $\mathfrak{B}_{12}$ we have to discuss the cases $f(r)\not\equiv0$ and $f(r)\equiv 0$ separately.
For the first case we can use
\begin{flalign}
\label{eqn:basisbh1}
 e_0=\partial_t~,~~e_1=f(r)\partial_r + c_2^0(r)\partial_t~,~~e_2=\partial_\zeta~,~~e_3=\partial_\phi~,
\end{flalign}
as a natural basis. For the second case, the twist vector fields can be reduced without loss of generality 
to $V_1=\kappa_1\partial_\phi$ and $V_2=c_2^0(r)\partial_t$, such that a natural basis would be
\begin{flalign}
\label{eqn:basisbh2}
 e_0=c_2^0(r)\partial_t~,~~e_1=\partial_r+ t c_2^0(r)^\prime/c_2^0(r) \partial_t~,~~e_2=\partial_\zeta~,~~e_3=\partial_\phi~,
\end{flalign}
where $c_2^0(r)^\prime$ denotes the derivative of $c_2^0(r)$.
It can be checked directly that (\ref{eqn:basisconditions}) is satisfied.

Next, we consider a metric field $g=\theta^a\otimes_\star\theta^b \star g_{ba}=\theta^a\otimes\theta^b g_{ba}$ in the natural basis.
Note that the $\star$-tensor product and $\star$-product in the expression above reduce to the undeformed products, since 
the twist acts trivially on the basis one-forms $\lbrace \theta^a\rbrace$. Furthermore, we have $g_{ab}=g_{ba}$.
 The inverse metric $g^{-1}=g^{ab}\star e_b\otimes_\star e_a=g^{ab} e_b\otimes e_a$ defined in~\cite{Aschieri:2005zs} satisfies
in the natural basis
\begin{flalign}
 g_{ab}\star g^{ca}=g^{ac}\star g_{ba} =\delta_b^c~,
\end{flalign}
i.\,e.~it is given by the $\star$-inverse matrix of $g_{ab}$. 

The $\star$-covariant derivative on tensor fields was defined in~\cite{Aschieri:2005zs}. Using the natural basis 
(\ref{eqn:basisconditions}) we obtain for its basis representation
\begin{flalign}
(\Nabla^\star_{e_c}\tau)_{b_1\dots b_l}^{a_1\dots a_n}=e_c(\tau^{a_1\dots a_n}_{b_1\dots b_l}) -\Gamma_{cb_1}^{~~\tilde b}\star\tau^{a_1\dots a_n}_{\tilde b\dots b_l} -\dots + \tau^{a_1\dots \tilde a}_{b_1\dots b_l}\star\Gamma_{c \tilde a}^{~~a_n}~.
\end{flalign}
Here $\Gamma_{ab}^{~~c}\star e_c:=\Nabla^\star_{e_a}e_b $ are the connection symbols and
$e_c(\cdot)$ is the vector field action on functions (Lie derivative).

In the natural basis the torsion tensor $\mathrm{T}=\theta^b\otimes\theta^a \mathrm{T}_{ab}^{~~c}\otimes e_c$ 
defined in~\cite{Aschieri:2005zs} reduces to
\begin{flalign}
 \mathrm{T}_{ab}^{~~c}=\Gamma_{ab}^{~~c} -\Gamma_{ba}^{~~c}~.
\end{flalign}
The metric compatible torsionfree connection is given by
\begin{flalign}
\label{eqn:connection}
  \Gamma_{ab}^{~~c}= \frac{1}{2}\bigl(e_a(g_{bd}) + e_b(g_{ad}) -e_d(g_{ab})\bigr)\star g^{cd}~.
\end{flalign}

In the natural basis the expression for the Riemann tensor $\mathrm{R}=\theta^c\otimes\theta^b\otimes\theta^a  \mathrm{R}_{abc}^{~~~d} \otimes e_d$
simplifies to
\begin{flalign}
\label{eqn:riemann}
 \mathrm{R}_{abc}^{~~~d}=e_a(\Gamma_{bc}^{~~d}) - e_b(\Gamma_{ac}^{~~d}) +\Gamma_{bc}^{~~e}\star\Gamma_{ae}^{~~d} - \Gamma_{ac}^{~~e}\star\Gamma_{be}^{~~d}~.
\end{flalign}
The Ricci tensor is given by $\mathrm{Ric}_{ab}=\mathrm{R}_{cab}^{~~~c}$ and the curvature scalar is given by
$\mathcal{R}=g^{ab}\star \mathrm{Ric}_{ba}$.

This leads to the NC Einstein equations proposed in~\cite{Aschieri:2005zs}
\begin{flalign}
\label{eqn:nceinstein}
G_{ab} := \mathrm{Ric}_{ab}-\frac{1}{2}g_{ab}\star\mathcal{R} = M_{pl}^{-2}T_{ab}~,
\end{flalign}
where we have introduced the Einstein tensor $G_{ab}$, the Planck mass $M_{pl}$ and a stress-energy tensor field $T_{ab}$. 
In this work we only need to assume that $T_{ab}$ is constructed from some (scalar) matter field $\phi$ and 
its covariant derivatives in a deformed covariant way.
We assume further that the stress-energy tensor is at least quadratic in the matter fields.

Note that in the natural basis all geometrical quantities defined above include only $\star$-products 
among the {\it coefficient functions} of tensor fields, and not among the basis vector fields and one-forms. 
Thus the formalism in the natural basis is more convenient for doing explicit calculations 
than the basis independent formalism of~\cite{Aschieri:2005zs,Aschieri:2006kc}.

To conclude this section we will briefly discuss possible issues with the NC Einstein equations (\ref{eqn:nceinstein}).
Firstly, it is not necessarily a real tensor field and secondly, the right hand side of the contracted 
second Bianchi identity
\begin{flalign}
\label{eqn:bianchi}
 g^{ba}\star (\Nabla^\star_{e_a} G)_{cb}=\frac{\Delta_c}{2}~,
\end{flalign}
does not vanish, where
\begin{multline}
 \Delta_c=g^{ba}\star\Bigl((\Nabla^\star_{e_a} \mathrm{Ric})_{cb}- (\Nabla^\star_{e_d}\mathrm{R})_{cab}^{~~~d}\Bigr)\\
-g^{ba}\star\Bigl(\starcom{\Gamma_{cd}^{~~\tilde d}}{\mathrm{R}_{\tilde d ab}^{~~~d}} - \starcom{\Gamma_{ad}^{~~\tilde d}}{\mathrm{R}_{\tilde d c b}^{~~~d} }  \Bigr) - (\Nabla^\star_{e_c} g^{-1})^{ba}\star \mathrm{Ric}_{ab}~.
\end{multline}

The first issue is not too dramatic and can in principle be solved by adding the
complex conjugate tensor, but the second issue in general leads to problems when coupling matter to gravity.
In this case the stress-energy tensor would have to satisfy
\begin{flalign}
 g^{ba}\star (\Nabla^\star_{e_a} T)_{cb}=\frac{\Delta_c}{2}~,
\end{flalign}
which is in general not compatible with sensible equations of motion for the matter fields for the case $\Delta_c\neq0$.

To solve this issue one could try to define a modified Einstein tensor $\tilde G_{ab}$, such that
\begin{flalign}
 g^{ba}\star (\Nabla^\star_{e_a} \tilde G)_{cb}=0~.
\end{flalign}
This could possibly be done for explicit problems by integrating the
right hand side of (\ref{eqn:bianchi}).  Fortunately, it will turn out
that for most of our models (\ref{eqn:models}) these problems do not
occur and we find $\tilde G_{ab}=G_{ab}$.  Because of this we postpone
the issue of modifying the Einstein tensor to a future work and only
discuss our well defined solutions in the following sections.

\section{Cosmological Solutions}
\label{sec:cosmo}

Firstly, we discuss exact cosmological solutions of the NC Einstein equations (\ref{eqn:nceinstein}).
For this we can use proposition 4 of~\cite{Ohl:2008tw} in order to find the right ansatz for the symmetry reduced 
metric field $g$.
This proposition tells us that a tensor field is invariant under the deformed action of the deformed symmetries, 
if and only if it is invariant under the undeformed action of the undeformed symmetries. The requirement for this proposition
was to use the so-called canonical embedding of the symmetry Lie algebra given by $\mathfrak{g}_\star=\mathfrak{g}$, which
 is fulfilled for our models defined in section~\ref{sec:models}. Therefore, we can make the ansatz
 $g=dx^\mu\otimes dx^\nu g_{\nu\mu}$ in the commutative coordinate basis, with
\begin{flalign}
 g_{\mu\nu}=\mathrm{diag}\Bigl(-1,A(t)^2,A(t)^2,A(t)^2\Bigr)_{\mu\nu}~,
\end{flalign}
and calculate the required coefficients $g_{ab}$ in the natural basis by solving
\begin{flalign}
 \theta^a\otimes\theta^b g_{ba}=dx^\mu\otimes dx^\nu \theta^a_\mu \theta^b_\nu g_{ba} = dx^\mu\otimes dx^\nu g_{\nu\mu}~,
\end{flalign}
 using the explicit expression of the natural basis vector fields (\ref{eqn:basiscosmo}).

It can be checked explicitly that for the choice $f_2=0$ in model $\mathfrak{C}_{22}$ (\ref{eqn:c22}) and 
the choice $V_1^0(t)\equiv 0$ in model $\mathfrak{C}_{32}$ (\ref{eqn:c32}) the NC connection symbols (\ref{eqn:connection}), the NC Riemann tensor
(\ref{eqn:riemann}) and finally the NC Einstein tensor (\ref{eqn:nceinstein}) receive no contributions in the deformation 
parameter $\lambda$, thus reducing to the undeformed counterparts. The reason for this is that for the restrictions
mentioned above we have one twist vector field $V_\alpha\in \mathfrak{g}$ and therefore deformed operations among
 symmetric tensors reduce to the undeformed ones, since $V_\alpha\in \mathfrak{g}$ annihilates the tensors due to invariance.

Since the NC stress-energy tensor of symmetric matter reduces to the undeformed tensor due to the same reasons,
these NC models are exactly solvable, iff the undeformed model is exactly solvable. Note that the reduction of the deformed
 symmetric tensors to the undeformed ones does not mean that our models are trivial. In particular, we will obtain in general
 a deformed dynamics for fluctuations on the symmetry reduced backgrounds, as well as a nontrivial coordinate algebra (\ref{eqn:commutators}).

Next, we will discuss physical implications of the nontrivial coordinate algebras of our models. Consider the model $\mathfrak{C}_{22}$
(\ref{eqn:c22}) with $f_2=0$ and for simplicity $V_1^0(t)\equiv1$. Then the coordinate algebra (\ref{eqn:a22}) 
reduces to the algebra of a quantum mechanical particle on the circle, i.\,e.
\begin{flalign}
 \com{\hat E}{\hat t}=\lambda \hat E~,
\end{flalign}
where we introduced the abstract operators $\hat t$ and $\hat E:=\widehat{\exp i\phi}$ and set $d=1$. This algebra previously appeared e.\,g.~in
 the context of noncommutative field theory~\cite{ncfield} and the noncommutative BTZ black hole~\cite{Dolan:2006hv}.
It is well known that the operator $\hat t$ can be represented as a differential operator acting on the Hilbert
space $L^2(S_1)$ of square integrable functions on the circle and the spectrum can be shown to be given by $\sigma(\hat t)=\lambda (\mathbb{Z}+\delta)$, where $\delta\in[0,1)$ labels unitary inequivalent representations. 
The spectrum should be interpreted as possible time eigenvalues.  
Thus our model with discrete time can be used to realize singularity avoidance
in cosmology.
Consider for example an
inflationary background with $A(t)=t^p$, where $p>1$ is a parameter. This so-called power-law inflation can be 
realized by coupling a scalar field with exponential potential to the geometry even in our NC model, since the symmetry reduced Riemannian geometry reduces to the undeformed one as explained above. Note that the scale factor goes to zero at the time $t=0$ and leads
to a singularity in the curvature scalar. But as we discussed above, the possible time eigenvalues are $\lambda (\mathbb{Z}+\delta)$,
 which does not include the time $t=0$ for $\delta\neq0$. 

Note that the discrete time and therewith possible singularity avoidance is a general feature of models with time-angle noncommutativity, even if $V_1^0(t)\neq \mathrm{const.}$. This can be seen by performing a local time reparametrization, such that
$V_1^0(t)\equiv1$, and pulling back the discrete spectrum of the time operator. The effect of this pullback is that the 
distance between the time eigenvalues will not be uniform in general.

For the more complicated solvable model $\mathfrak{C}_{32}$ (\ref{eqn:c32}) with $V_1^0(t)\equiv0$ we obtain time-angle and angle-radius
 noncommutativity. We set without loss of generality the parameter $d=f_2=1$. Firstly, we choose $V_2^0(t)\equiv0$ leading to a pure angle-radius noncommutativity. The algebra (\ref{eqn:a32})
becomes in this case
\begin{flalign}
 \hat E \hat r=e^{-\lambda} ~\hat r \hat E~.
\end{flalign}
This algebra can be represented on the Hilbert space $L^2(S_1)$ as
\begin{flalign}
 \hat E = \exp i\phi~,\quad \hat r = \Lambda \exp{\bigl(-i\lambda \partial_\phi\bigr)}~,
\end{flalign}
leading to the spectrum $\sigma(\hat r)=\Lambda \exp\bigl(\lambda (\mathbb{Z}+\delta)\bigr)$, where $\delta\in[0,1)$ is again a parameter labeling unitary inequivalent representations and $\Lambda$ is some length scale. 
Therefore the radius becomes discrete. It would be natural to choose
for $\Lambda$ a physical length scale (e.\,g.~the Hubble length)
in order to avoid a growth of the eigenvalue spacings with time in an expanding universe.
Note that this model describes a kind of ``condensating geometry'' around the origin $r=0$, since the shells of 
constant $r$ accumulate at this point. It remains to be shown for which values of $\lambda$ and $\Lambda$ this is consistent 
with experimental data from the cosmic microwave background (CMB).

Secondly, we choose for simplicity $V_2^0(t)\equiv1$ in (\ref{eqn:a32}). In the case of nonconstant $V_2^0(t)$ 
we can in principle pull back the spectrum as discussed before. We obtain the abstract algebra
\begin{flalign}
\label{eqn:alg32}
 \com{\hat t}{\hat E}= \lambda \hat E~,\quad \hat E \hat r= e^{-\lambda} ~\hat r \hat E~. 
\end{flalign}
Furthermore, we use the representation on $L^2(S_1)$
\begin{flalign}
 \hat E = \exp i\phi~,\quad\hat t=\tau \hat 1 - i\lambda \partial_\phi~,\quad \hat r = \Lambda \exp{\bigl(-i\lambda \partial_\phi\bigr)}~.
\end{flalign}
Note that we had to introduce a real parameter $\tau\in\mathbb{R}$ and the identity operator $\hat 1$ 
in order to cover the whole spacetime.

The last cosmological model we want to briefly discuss is the isotropic model $\mathfrak{C}_{22}$ with $d=0$ (\ref{eqn:c22}).
It turns out that both $V_\alpha\not\in\mathfrak{g}$, for the
Riemannian geometry not to reduce to
the undeformed one. Thus we expect corrections in $\lambda$ to the NC Einstein equations (\ref{eqn:nceinstein}) and its solutions.
Since it is not yet clear how to formulate consistent NC Einstein equations coupled to matter, we postpone the investigation of 
these corrections to a future work and give here only one special exact solution of this model.

Consider the (undeformed) de Sitter space given by $A(t)=\exp H t$, where $H$ is the Hubble parameter.
It turns out that all $\star$-products entering the deformed geometrical quantities (see section~\ref{sec:basics})
reduce to the undeformed ones, if $V_1^0(t)\equiv1$. Thus the undeformed de Sitter space solves NC Einstein equations (\ref{eqn:nceinstein}),
 or possible modifications of it, for this particular choice of twist and an undeformed cosmological constant. 
 Note that in contrast to the solutions above, we required the explicit form of the scale factor $A(t)$. 

This shows that we can construct at least one
 exact solution of the isotropic model. The general case still requires further investigation.

\section{Black Hole Solutions}
\label{sec:blackhole}

For the black hole model (\ref{eqn:b12}) we use again proposition 4 of~\cite{Ohl:2008tw} and make the ansatz
\begin{flalign}
\label{eqn:bhmetric}
 g_{\mu\nu}=\mathrm{diag}\Bigl(-Q(r),S(r),r^2,(r\sin\zeta)^2  \Bigr)_{\mu\nu}
\end{flalign}
for the metric field $g=dx^\mu\otimes dx^\nu g_{\nu\mu}$ in the commutative spherical coordinate basis. The metric in the 
natural basis can be calculated using (\ref{eqn:basisbh1}) or (\ref{eqn:basisbh2}), respectively. 
Concerning the solution of the NC Einstein equations we are in a comfortable position, since 
we have $V_1\in\mathfrak{g}$, which means that the symmetry reduced Riemannian geometry
 reduces to the undeformed one. This leads in the exterior of our NC black hole to the metric (\ref{eqn:bhmetric}) with
\begin{flalign}
 Q(r)=S(r)^{-1} = 1-\frac{r_s}{r}~,
\end{flalign}
where $r_s$ is the Schwarzschild radius. As in the case of the cosmological models, the reduction of the symmetry reduced tensor fields
to the undeformed counterparts does not mean that our models are trivial. Fluctuations (e.\,g.~Hawking radiation), as well as 
the coordinate algebras will in general receive distinct NC effects.

Taking a look at the coordinate algebra of the black hole (\ref{eqn:a12}), we observe that it includes in particular the
algebra of a quantum mechanical particle on the circle for time and angle variable, if we choose $c_2^0(r)\equiv0$ and 
$f(r)\equiv 0$. This leads to discrete times.
Another simple choice is $c_1^0=\kappa_2=0$, $c_2^0(r)\equiv0$, $\kappa_1=1$ and $f(r)=r$. The radius spectrum in this case is 
$\sigma(\hat r)=\Lambda\exp\bigl(\lambda(\mathbb{Z}+\delta) \bigr)$, describing a fine grained geometry around the black hole.
The phenomenological problem with this model is that the spacings between the radius eigenvalues grow exponentially in $r$.
This can be fixed by considering a modified twist like e.\,g.~$c_1^0=\kappa_2=0$, $c_2^0(r)\equiv0$, $\kappa_1=1$ and $f(r)=\tanh \frac{r}{\Lambda}$, where $\Lambda$ is some length scale.
The essential modification is to choose a bounded $f(r)$. 
Consider the coordinate change $r\to \eta=\log \sinh (\frac{r}{\Lambda})$, then the
algebra (\ref{eqn:a12}) in terms of $\eta$ becomes
\begin{flalign}
 \com{\hat E}{\hat \eta}=-\frac{\lambda}{\Lambda} \hat E~,
\end{flalign}
leading to the spectrum $\sigma(\hat \eta)=\frac{\lambda}{\Lambda} \bigl(\mathbb{Z}+\delta\bigr)$.
The spectrum of $\hat r$ is then given by $\sigma(\hat r) = \Lambda~ \mathrm{arcsinh}\exp\bigl(\frac{\lambda}{\Lambda}(\mathbb{Z}+\delta)\bigr) $. This spectrum approaches constant spacings
 between the eigenvalues for large $r$.

We omit a deeper discussion of further possible models, since our main
purpose was to present the very explicit and simple models shown
above. We conclude this section with one remark. Our class of black
hole models (\ref{eqn:b12}) is related to the NC black hole models
found earlier by Schupp and Solodukhin~\cite{schuppbh}. They also
found that the symmetry reduced dynamics reduces to the undeformed one
for their black hole models. In addition, they constructed models
based on a projective twist, that is not contained in the RJS-class,
which exhibit discrete radius eigenvalues as well.

\section{Conclusions and Outlook}
\label{sec:conc}

In this paper we have constructed exact cosmological and black hole
solutions of the noncommutative gravity theory proposed
in~\cite{Aschieri:2005yw,Aschieri:2005zs}. In particular we have
obtained FRW models in which the physical scale of noncommutativity is
not growing with time, as it would be for the most simple Moyal-Weyl
deformation.  Some of our models possess interesting physical
features, such as a discrete time spectrum for the case of FRW models
and a discrete radius spectrum for the black hole.  Furthermore, we
have found that the most attractive cosmological model, deformed by an
isotropic twist, solves the NC Einstein equations in presence of a
cosmological constant.  We also found that in particular for the
isotropic twist FRW model, noncommutativity can in general influence
the dynamics of the symmetry reduced sector. It will therefore be
interesting to study, if we can use noncommutativity in order to drive
inflation.  In order to study these effects, one should modify the NC
Einstein tensor as proposed in section~\ref{sec:basics} or use the
recently proposed NC vielbein gravity~\cite{vielbein} in order to
couple matter and geometry properly.

In future work~\cite{Ohl/Schenkel:fluct} we will construct scalar field fluctuations on curved NC backgrounds 
in a twisted covariant setting.
This would also include the twist deformation of the Poisson algebra of field observables, as it was done in~\cite{poisson} for the case of
the Moyal deformed Minkowski spacetime. Using this formalism, we will study NC modifications of the cosmic microwave background (CMB) 
and possibly also Hawking radiation. It will also be interesting to compare our approach with existing results on NC
effects in the CMB, obtained in different settings~\cite{cmb}.

Recently the physics of noncommutative Kerr black hole was
studied~\cite{NCKerr} in the framework of~\cite{Aschieri:2005yw}.  It
should be fruitful to study their results in our approach.

\acknowledgments
The authors thank Peter Schupp for valuable discussions and comments.
Furthermore, AS thanks Christoph Uhlemann and Julian Adamek for some comments.  
This research is supported by Deutsche
Forschungsgemeinschaft through the Research Training Group GRK\,1147
\textit{Theoretical Astrophysics and Particle Physics}.


\end{document}